\documentstyle[11pt,rtrpp4]{article}

\def\alwaysmath#1{\ifmmode{#1}\else{$#1$}\fi}
\def\mhyph{\rm -}
\def\msun{\alwaysmath{\,M_\odot}}

\def\morethan#1{{ et al.}}  \def\pc{{\,\rm pc}}
\def\bsshb{{F_{\scriptscriptstyle\rm HB}^{\scriptscriptstyle\rm BSS}}}
\newcommand\hst{{\it HST}} 
\newcommand\etal{et~al.}  
\slugcomment{\footnotesize In press to ApJ.
}

\lefthead{Ferraro et al.}
\righthead{BSS: The Spectacular Population in M80}

\begin{document}

 \title{Blue Straggler Stars: The Spectacular Population 
in M80\footnote{Based on
observations with the NASA/ESA {\it Hubble Space Telescope}, obtained at
the Space Telescope Science Institute, which is operated by AURA, Inc.,
under NASA contract NAS5-26555}}

\author{Francesco R. Ferraro\altaffilmark{2,3},  
Barbara Paltrinieri\altaffilmark{4},  
Robert T. Rood\altaffilmark{5},
Ben Dorman\altaffilmark{6}
}
\altaffiltext{2}{European Southern Observatory, Karl Schwarzschild Strasse 2,
D-85748 Garching bei M\"unchen, Germany}
\altaffiltext{3}{Osservatorio Astronomico di Bologna, Via Ranzani 1, 40127 
Bologna, ITALY}
 \altaffiltext{4}{Instituto di Astronomia---Universit\`a  ``La Sapienza,''
Piazzale Aldo Moro 5, 00185 Roma, ITALY}
\altaffiltext{5}{Astronomy Dept, University of Virginia,
	P.O.Box 3818, Charlottesville, VA 22903-0818}
\altaffiltext{6}{Raytheon STX \& Laboratory for Astronomy \& Solar Physics,
Code 681, NASA/GSFC,	Greenbelt MD 20771}

\begin{abstract}

Using {\it HST-WFPC2} observations in two ultraviolet (UV) filters
(F225W and F336W) of the central region of the high density Galactic
Globular cluster (GGC) M80 we have identified 305 Blue Straggler Stars
(BSS) which represents the largest and most concentrated population of
BSS ever observed in a GGC. We also identify the largest, clean sample
of evolved BSS yet found. The high stellar density alone cannot
explain the BSS, and we suggest that in M80 we are witnessing a
transient dynamical state, during which stellar interactions are
delaying the core-collapse process leading to an exceptionally large
population of {\it collisional}-BSS.

\end{abstract}


\section{Introduction}

Blue straggler stars (BSS) where first observed in the 1950's
(\cite{sandage53}) in the Galactic globular cluster (GGC) M3. In the
color-magnitude diagram they formed a sparsely populated sequence
extending to higher luminosities than the turn-off point of normal
hydrogen burning main-sequence stars. Superficially they looked like a
population of younger stars, more massive than the turn-off stars, in
an old star cluster. Since there is no other indication of star
formation after the burst which formed the bulk of the cluster stars,
two mechanisms for making BSS are favored. First is the merger of two
stars in a primordial binary system, where ``primordial'' refers to
binaries formed when the cluster formed. Second are collisions in
regions of very high stellar density (\cite{hills76}, \cite{fpfc92},
\cite{bss2pop}, \cite{bss95}, \cite{bailynaraa}, \cite{mh97}). These
collisional-BSS include several classes of objects: direct collisions
producing a more massive star; collisions which harden primordial
binaries until the point of merger; and binaries produced in
collisions which later merge. The dense cores of globular clusters
were obvious targets for observations required to refine our
understanding of BSS. Indeed, more than 20 years ago Hills \& Day (1976)
suggested searching the core of M80 for collisional BSS. However, only
with the advent of the Hubble Space Telescope ({\it HST}) could such
observations be made (\cite{paresce47tuc}, \cite{m15fp}, \cite{m3bss},
\cite{dsm5}, \cite{guham30}).

The BSS population, especially collisional-BSS, can serve as a
diagnostic for the dynamical evolution of GGCs.  Because of
gravitational interactions between cluster stars, GGCs evolve
dynamically on time scales generally smaller than their ages. For
example, the first manifestation of dynamical process within a GGC is
that in the inner part of a GGC more massive stars (or binaries)
should settle toward the center.  Beyond this, more dramatic dynamical
phases can happen during the cluster's lifetime.  Stars with
velocities above the escape velocity continuously evaporate, and
phenomena such as Galactic tidal stripping remove stars from the
outer regions of the cluster and induce substantial changes in the
structure of the cluster itself.  As a GGC adjusts to the loss of
stars, the cluster core must contract. Under some circumstances this
process can run away leading to a possibly catastrophic
``core-collapse.'' About 15\% of the GGC population show evidence for
this phenomenon.  Binaries are thought to play a fundamental role in
the core collapse: binary-binary collisions could in fact be effective
in halting (or, more probably, delaying) the collapse of the core
avoiding infinite central density.

This time of enhanced binary interactions as the cluster fights off core
collapse could well correspond to a period of unusually large BSS
production.  By the end of this phase most of the binaries in the core
will be destroyed by close encounters; the survivors will become highly
hardened (i.e., tightly bound), producing most of the additional
collisional-BSS.

\section{Observations}

To search for BSS (and other blue objects) we have used the {\it Wide
Field Planetary Camera} (WFPC2) of \hst\ to obtain ultraviolet and
visible images of the central region of the high density cluster M80
(NGC~6093). Both the high angular resolution and UV sensitivity of
\hst\ are essential to identify these UV-bright objects among the much
more luminous red giants in the cluster (\cite{m3bss}). The images
were obtained on 5--6 April 1996 (GO-5903, PI: F.R. Ferraro) with the
WFPC2 F160BW (far-UV), F255W (mid-UV), F336W ($U$), and F555W (visible
or $V$) filters.  The Planetary Camera (PC, which has the highest
resolution $\sim 0\farcs{046}/{\rm pixel}$) was roughly
centered on the cluster center while the Wide Field (WF) cameras (at
lower resolution $\sim 0\farcs{1}/{\rm pixel}$) sampled the
surrounding outer regions. The BSS identifications are based on
$4\times600\,$s exposures in $U$ and $4\times300\,$s exposures in
F255W.  The WFPC2 frames were processed through the standard HST-WFPC
pipeline and photometry was obtained as outlined in our study of BSS
in M3 (\cite{m3bss}). Figure~\ref{image} shows the advantages of using
UV images to search for BSS: in the center of the $V$ image the light
from the bright red giant branch (RGB) stars blends together. In the
UV image the brightest objects are horizontal branch (HB) stars and
BSS; there is little blending even at the center.

\section{Results}

The Ultraviolet Color Magnitude Diagram (CMD) in the
($m_{255},m_{255}-U$) plane for more than 13,000 stars identified in
the HST field of view, is presented in Figure~\ref{uvcmd}.  The large
population of BSS defines a narrow nearly-vertical sequence spanning
$\sim 3 $ mag in $m_{255}$. They are clearly separable from the cooler
and fainter Turn-Off and subgiant branch (SGB) stars.  However, as
already discussed in previous papers (see \cite{bss95}) one of the
major problem in defining homogeneous samples of BSS is the {\it
operative} definition of the faint edge of the BSS population. This is
true even in UV-CMDs (see for example \cite{m3bss}), since generally
the BSS sequence merges smoothly into the $\rm MS+TO$ region without
showing any gap or discontinuity.  In selecting the BSS sample here we
have adopted the same criteria we used in M3, which was recently
observed (\cite{m3bss}) with the same technique and set-up used here.
In order to assure the same BSS limiting absolute magnitude for M80 as
we adopted in M3, we aligned the two $(m_{255},~m_{255}-U)$ CMDs,
using the bright portion of the HB as {\it normalization} region. The
shift in magnitude required to align the two CMDs is $\delta m_{255} =
1.15$. The resulting fainter boundary of the BSS sequence in M80 is
$m_{255}=20.55$. Adopting this figure M80 turns to have a
spectacularly large population of BSS---305 candidates have been found
in the WFPC2 field of view.

Ferraro \etal\ (1997a) split the M3 BSS into {\it bright} and {\it faint}
subsamples. The analogous division in  M80 is at
$m_{255}=20.15$. M80 has (1) 129 bright BSS with $m_{255}<20.15$
and (2) 176 faint BSS with $20.15<m_{255}<20.55$.

The BSS region in the CMD is better shown in panel (b) of
Figure~\ref{uvcmd} where the total sample of BSS is plotted as big dots.
Note that the limiting magnitude for the faint BSS is at the ``error
envelope'' of the main sequence region on the CMD. By examining the
adjacent regions of the CMD we estimate that there at most a few MS
stars misidentified as BSS. In addition the faint BSS and bright BSS
have almost identical radial distributions while that of the MS stars is much
less centrally concentrated, similar to that of the RGB+HB stars (see
Fig.~\ref{cumdist} below). This again suggests at most a very minor
contamination of the faint BSS sample.

Table 1 lists the BSS candidates: the first column is the number, then in
columns 2--5 we report the identification number, $m_{255}$ and $U$
magnitudes and the coordinates $(X,~Y)$, respectively.  The coordinates are
referred to an arbitrary system and are expressed in {\it ground-based}
pixel units (1 pixel $=0\farcs 35$), after a rotation and translation to
match the complementary ground-based observations (see below).
 
While not obvious from Figure~\ref{image}, Figure 3 clearly shows that
the BSS (heavy solid line) are far more concentrated towards the
cluster center than either the HB or RGB stars. (The dashed line shows
the combined distribution of the HB+RGB which are individually quite
similar).  Half of the BSS population is within $8\arcsec$ from the
cluster center, compared to only $\sim 20\%$ of the HB or RGB in the
same region.  The Kolmogorov-Smirnov test applied to the two
distributions shows that the probability of drawing the two
populations from the same distribution is very small, $\sim 10^{-4}$.
This result is consistent with the scenario that BSS are much more
massive population than normal HB, RGB stars. A recent direct
spectroscopic mass measured for a BSS in the core of the GGC 47~Tuc
(\cite{47tucmbss}) also indicates  a higher mass
for that star.

Extensive {\it artificial star} tests have been performed to estimate the
degree of completeness of the detected BSS population.  The completeness
level is $>80\%$ at the faint edge of the bright sample and $\sim 72\%$ at
the faintest magnitude limit. From these results we estimate that the
{\it true} number of BSS in M80 could be as large as $\sim 400$.

The number of BSS in M80 is huge. The previous record number was in M3
which has a population of $\sim 170$ BSS (about half of the population in
M80) in the WFPC2 field of view (\cite{m3bss}).  A quantitative comparison
requires that the BSS number be normalized to account for the size of the
total population.  This is done with an appropriate specific frequency:

\begin{displaymath}
\bsshb = {{N_{\scriptscriptstyle\rm BSS}} \over
{N_{\scriptscriptstyle\rm HB}}}
\end{displaymath}

\noindent where $N_{\scriptscriptstyle\rm BSS}$ is the number of BSS
and $N_{\scriptscriptstyle\rm HB}$ is the number of HB stars in the
same area. This ratio can be easily computed in the UV-CMDs since the
HB population is quite bright and the sequence well defined.  The
specific frequency of BSS in M80 turns to be $\sim 1$.  In other
clusters with similar mass, M3, M13 and M92, which have been observed
with similar technique by our group we find substantially lower values
ranging from $\bsshb \sim 0.17$ for M13 up to 0.55 and $0.67$ for M92
and M3.  Moreover, considering only the field of view of the PC, the
specific frequency of BSS in M80 rises to $\sim 1.7$, i.e., the BSS
are almost twice as abundant as the HB stars.

 Several other clusters have recently been surveyed with the WFPC2
covering a region comparable with that of our
observations. The somewhat less massive cluster M30 has a population
of 48 BSS and a specific frequency $\bsshb =0.49$ (\cite{guham30}).
While not optimal for BSS searches, the survey of Sosin \etal\ (1997) can
give a rough indication of the central BSS population.  The clusters
with the largest BSS population are NGC~6388 and NGC~2808 each with
$\sim 100$ BSS. These clusters are each about a factor 4 more massive
than M80 but still contain only a fraction ($\sim 0.3$) of the BSS
population found in M80. The corresponding specific frequencies of BSS
would be about 0.1 that of M80. Either in terms of number or specific
frequency M80 becomes the Galactic BSS record holder.

\section{Discussion}

One might speculate that the BSS in M80 are produced by an anomalously
large population of primordial binaries.  If so, some of these
binaries should be detectable outside the cluster core in the form
primordial-binary-merger BSS, such as those found in the outer region
of M3 (\cite{ph94,bss2pop,m3bss}). However, recent CMDs of the outer
parts of M80 (\cite{bcsspf98,alcainom80}) give no indication for a
large {\it primeval} population comparable to that found in M3.  Given
this we turn to the structural characteristics of M80 for an
explanation.

M80 is much more centrally condensed than M3, M92, and M13, a factor
that might promote the production of collisional-BSS. Can that factor
alone account for the BSS population? We suspect not, because the BSS
population in M80 is also large compared with other clusters with high
central density. For example, the central part of 47~Tuc $\log\rho_0
\sim 5.1 \msun\pc^{-3}$ compared to $5.4\msun\pc^{-3}$ for M80 and in
contrast to $3.5\msun\pc^{-3}$ for M3 (\cite{pm93}).  Figure 1 of
Sosin \etal\ (1997) shows that 47 Tuc does not have a large population of
BSS---no more than 50 BSS can be counted. Likewise, NGC~2808 and
NGC~6388 have densities of $\log\rho_0 \sim 4.9~{\rm and~} 5.7
\msun\pc^{-3}$ respectively and relatively modest BSS populations.

Since high density cannot account for the large number of BSS in M80
perhaps they arise from its dynamical state.  M80 has one of the
highest central densities ($\log\rho_0 \sim 5.4 \msun\pc^{-3}$) of any
GGC which has shown no previous evidence for having undergone core
collapse (\cite{djor93}). Generally GGCs are considered core-collapsed
or not depending on how well their radial distribution of stars is fit
by King Models (\cite{king66}). These models are characterized by two
parameters, the core radius, $r_c$, and the tidal radius, $r_t$, or,
alternatively, the concentration, $c= \log(r_t/r_c)$.  Our data
supplemented with ground-based observations (\cite{bcsspf98}) for
$r>85\arcsec$ provides the best such test to date for M80.

To determine $r_c$ and $c$ we first determined the gravity center
$C_{\rm grav}$ following the procedure of Montegriffo \etal\ (1995).  We
computed $C_{\rm grav}$ by simply averaging the $X$ and $Y$ coordinates (in
the local system) of stars lying in the PC camera, and then transforming
them to the absolute system.  $C_{\rm grav}$ is located at pixel
$(503\pm5,~418\pm5)$ in our PC image; this  corresponds to:
$\alpha_{\rm J2000} = 16^{\rm h}\, 17^{\rm m}\, 02\fs 29, 
\delta_{J2000} = -22\arcdeg\, 58\arcmin\, 32\farcs 38$ which is $\sim 4\arcsec$
NW of the center reported in the Djorgovski (1993) compilation.  The
$C_{\rm grav}$ is at pixel $(676,~647)$ in the ground-based 
coordinate system used in Table~1.

The density
profile with respect to the measured gravity center $C_{\rm grav}$
 is shown in
Figure~\ref{kingmods}.  It was derived using the standard technique
(\cite{d88}) for all stars with $V<19.5$. A King model with the most
recent values (\cite{trageretal93}), $r_c=9\arcsec$ and $c=1.95$, does
not reproduce the observed density profile for $r<8\arcsec$, however a
King model with a smaller $r_c=6\farcs{5}$ and essentially the same
$c=2.0$ fits the data reasonably well as seen in
Figure~\ref{kingmods}.

Meylan \& Heggie (1997) warn that it can be difficult to differentiate
the dynamical (pre- in- or post-collapse) phase of a GC on the basis
of the shape of the density profile. However, they suggest, as a rule
of thumb, that {\it ``any GC with a concentration $c\sim2.0-2.5$ may
be considered as collapsed or on the verge of collapsing or just
beyond.''}  Thus, while the good fit to the King model suggests that
M80 has not yet completed core-collapse, the value of $c$ is
consistent with the suggestion that M80 is on the verge of
collapse. The other piece of information we can bring to bear is the
anomalously large BSS population. Two PCC clusters have been observed
deep enough and with appropriate filters that we have a reasonable
estimate of their central BSS populations. Neither of these, 47~Tuc
(\cite{sosin-ase}) and M30 (\cite{guham30}), has a BSS frequency close
to that of M80. Thus we see that being in a PCC state can not
explain the BSS population of M80. 

The most plausible hypothesis at this point is that the BSS arise from
the core collapse process. It is commonly thought that binaries play
an important role on the core collapse (\cite{hut92,mh97}) with the
formation of binaries delaying and eventually halting the
collapse. With its high central density M80 is probably trying very
hard to undergo core collapse but binaries are forming and preventing
this from happening. A large population of collision-BSS should exist
during this time and slightly beyond (until the BSS begin to die off).

This scenario is fully compatible with dynamical evolution times:
following Meylan \& Heggie (1997), without including binary
formation the entire evolution time ($t_{ce}$) of the core is
$t_{ce}\sim 16 t_{rh}(0)$ where $t_{rh}(0)$ is the initial half mass
relaxation time.  Using values from Djorgovski (1993), we obtain for
M80 $t_{ce}\sim 4 \times 10^8$, which is 30 times smaller than the
cluster's age.

\section{The Evolved BSS}

With such a large population of BSS we might expect to find a
significant population of evolved BSS (E-BSS).  Renzini \& Fusi Pecci
(1988) suggested searching for E-BSS during their core helium burning
phase since they should appear to be redder and brighter than {\it
normal} HB stars.  Following this prescription Fusi Pecci et
al. (1992) identified a few E-BSS candidates in several clusters with
predominantly blue HBs where the likelihood of confusing E-BSS stars
with true HB or evolved HB stars was minimized. Because of the small
numbers there always the possibility that some or even most of these
candidate E-BSS were due to field contamination. Near cluster centers
field contamination should be less of a problem. In our \hst\ study of
M3 we identified a sample of E-BSS candidates (see Ferraro et al 1997)
and argued that the radial distribution of E-BSS was similar to that
of the BSS. M80 offers some advantages over M3 in searching for E-BSS:
1) it has a very blue HB so there should be less confusion between red
HB stars and E-BSS; 2) it has a larger number of BSS; 3) we have
identical photometry for M13 which has a very similar BHB to M80
coupled with a much smaller number of BSS---the E-BSS region of the
CMD of M80 should have a substantially larger number of stars than
that of M13.  In Figure 3a we show a zoomed $(U,~U-V)$ CMD of the HB
region.  The expected location for E-BSS has been indicated as a box;
19 E-BSS (plotted as large filled circles) lie in the box. There are
only 5 E-BSS in the same part of the CMD of M13.  $V$ and $U$
magnitudes and position for the E-BSS found in M80 are listed in Table
2.

In the case of M80 it is very unlikely that the E-BSS population is due to
background field contamination.  In fact, most (15) of the E-BSS have been
found in the PC field of view, while only 4 E-BSS lie in the most external
WFs.  A estimate of the expected field contamination can be computed
adopting the star counts listed by Ratnatunga \& Bahcall (1985). Following
their model, $\sim 0.6$ star per square arcmin is expected in a section of
the CMD which is {\it twice} the size of the region used to isolate the
E-BSS population. (The E-BSS span less than 1 magnitude in $V$, while the
Ratnatunga \& Bahcall (1985) counts are listed for 2 mag-wide bins.)  The
expected number of field stars is 0 in the PC field of view and 1.6 stars
in the global field of view of the three WF cameras.  For this reason we
can reasonably conclude that the region of the CMD used to select E-BSS
candidate is essentially unaffected by field contamination.

The cumulative radial distribution of the E-BSS stars is  shown
(as dotted line) in Figure 3.
The E-BSS cumulative distribution is quite similar to the BSS
distribution and significantly different from that of the HB-RGB.  A
Kolmogorov-Smirnov test shows that the probability that the E-BSS and
BSS population has been extracted from the same distribution
is $\sim 67\%$ while the probability that the E-BSS and the
RGB-HB population have the  same distribution is
only $\sim 1.6 \%$. This result confirms the expectation that the 
E-BSS share the same distribution of the BSS and they are
both a more massive population than the bulk of the
cluster stars. It further strengthens the case that field contamination is
negligible.

Earlier studies (\cite{fpfc92,m3bss}) have suggested that the ratio of
bright BSS (b-BSS) to E-BSS is $N_{\rm b\mhyph BSS}/N_{\rm E\mhyph BSS} \approx
6.5$. For M80 the number of b-BSS (defined as in \cite{m3bss}) is
$N_{\rm b\mhyph  BSS} = 129$, and we find $N_{\rm b\mhyph BSS}/N_{\rm
E\mhyph BSS} = 6.8$ fully consistent with earlier
studies. Because both our BSS and E-BSS samples are so cleanly defined
the ratio of the total number of BSS to E-BSS, $N_{\rm BSS} /N_{\rm
E\mhyph BSS} \sim 16$, should be useful in testing lifetimes of
BSS models.

\section{Conclusions}

The emerging scenario for BSS is complex. All GGCs which have been
properly surveyed have some BSS, so BSS must be considered as a normal
component of GGC population. BSS are found in diverse environments and
are probably formed by both merging primordial binaries and stellar
collisions. Some intermediate-low density clusters have only a few BSS
(M13) while similar clusters (M3, M92) have many more. This may arise
from the fact that the initial population of binaries in clusters like
M13 is small. The relatively large population of BSS in the exterior
of M3 (\cite{m3bss}) in contrast to the absence of BSS in the exterior
of M13 (\cite{paltrinm13}) supports the notion of very different
primordial binary populations. 

The densest
cluster cores have significant but highly variable BSS
populations (see the discussion in Ferraro, Bellazzini \& Fusi Pecci 1995).
In particular the post-core-collapse clusters 47~Tuc and M30 have
significantly smaller BSS populations than M80.

 We suggest that exceptional population in M80 arises because we have
caught a cluster at a critical phase in its dynamical evolution. This
effect could be enhanced by a large fraction of primordial binaries,
but there is no indication for this in the form a large BSS population
in the outer cluster (\cite{bcsspf98,alcainom80}).  More
information is needed before a definitive conclusion can be reached. A
search for other indications of a high frequency of stellar
multiplicity in M80, such as a broadening of the main sequence, would
also be very useful.  Also, further study of the velocity distribution
would be important to clarify the dynamical state of the cluster
(\cite{mh97}).  Core collapse is one of the most spectacular phenomena
in nature. It is important to confirm whether we have caught M80
during the period when the stellar interactions are delaying the
collapse of the core (and producing BSS).

\acknowledgments
This research was partially financed by the Agenzia
Spaziale Italiana (ASI). FRF acknowledges the  MURST  financial support to  
the project {\it Stellar Evolution} and the {\it ESO Visitor Program}
for its hospitality.
RTR \& BD are supported in part by the NASA Long Term Space Astrophysics
grant NAG 5-6403 and STScI/NASA grants  GO-6607, 6804.

\newpage
\hoffset = -17mm
\begin{deluxetable}{ccccccccccccc}
\tablewidth{19truecm}
\label{lm}
\tablecaption{The BSS population in M80.}
\tablehead{
\colhead{ $Name$} &
\colhead{ $Id$} &
\colhead{$m_{255}$} &
\colhead{$U$} &
\colhead{$X$} &
\colhead{$Y$} & 
\colhead{} &
\colhead{ $Name$} &
\colhead{ $Id$} &
\colhead{$m_{255}$} &
\colhead{$U$} &
\colhead{$X$} &
\colhead{$Y$} }
\startdata
BSS1&  13166 &   18.272 &   17.946 &   673.806 &  609.805 &  
&BSS41&  14383 &   19.726 &   19.053 &   677.829 &  666.232  \nl
BSS2&  13676 &   18.661 &   18.248 &   678.657 &  655.898 &  &BSS42&  12212 &   19.740 &   18.739 &   695.089 &  658.973  \nl
BSS3&  43933 &   18.662 &   18.210 &   650.530 &  605.157 &  &BSS43&  13221 &   19.766 &   19.250 &   652.808 &  632.693  \nl
BSS4&  10338 &   18.751 &   18.207 &   614.898 &  649.000 &  &BSS44&  13547 &   19.768 &   19.143 &   670.100 &  649.470  \nl
BSS5&  14551 &   18.753 &   18.365 &   669.207 &  637.937 &  &BSS45&  11923 &   19.788 &   19.109 &   679.463 &  660.599  \nl
BSS6&  14973 &   18.846 &   18.384 &   683.438 &  648.577 &  &BSS46&  13813 &   19.806 &   19.137 &   660.437 &  679.883  \nl
BSS7&  43603 &   18.867 &   18.392 &   652.347 &  601.041 &  &BSS47&  13497 &   19.808 &   19.234 &   685.882 &  634.075  \nl
BSS8&  20052 &   18.942 &   18.527 &   614.882 &  667.532 &  &BSS48&  14261 &   19.809 &   18.634 &   674.965 &  647.004  \nl
BSS9&  11193 &   18.964 &   18.694 &   673.794 &  640.225 &  &BSS49&  15206 &   19.810 &   18.926 &   673.290 &  643.834  \nl
BSS10&  13786 &   19.032 &   18.554 &   709.550 &  645.030 &  &BSS50&  14365 &   19.811 &   18.735 &   685.082 &  658.333  \nl
BSS11&  14804 &   19.040 &   18.556 &   669.121 &  630.334 &  &BSS51&  13238 &   19.817 &   18.864 &   689.182 &  610.791  \nl
BSS12&  13180 &   19.062 &   18.307 &   673.960 &  612.312 &  &BSS52&  14584 &   19.834 &   18.807 &   668.031 &  653.008  \nl
BSS13&  15327 &   19.152 &   18.577 &   680.614 &  647.182 &  &BSS53&  11160 &   19.845 &   19.167 &   656.767 &  649.971  \nl
BSS14&  11215 &   19.185 &   18.213 &   663.463 &  647.503 &  &BSS54&  15180 &   19.860 &   18.859 &   694.186 &  653.441  \nl
BSS15&  15329 &   19.224 &   18.618 &   680.730 &  647.575 &  &BSS55&  13380 &   19.862 &   19.371 &   667.483 &  638.571  \nl
BSS16&  15356 &   19.262 &   18.398 &   680.337 &  650.271 &  &BSS56&  15111 &   19.867 &   18.967 &   687.050 &  645.576  \nl
BSS17&  14208 &   19.263 &   18.614 &   675.750 &  638.118 &  &BSS57&  12270 &   19.869 &   19.079 &   708.024 &  652.182  \nl
BSS18&  13834 &   19.293 &   18.736 &   706.016 &  652.431 &  &BSS58&  11316 &   19.874 &   19.153 &   698.037 &  628.225  \nl
BSS19&  11184 &   19.313 &   18.719 &   697.225 &  624.590 &  &BSS59&  15217 &   19.876 &   19.148 &   675.268 &  644.626  \nl
BSS20&  11610 &   19.315 &   18.674 &   674.402 &  653.767 &  &BSS60&  20127 &   19.888 &   19.091 &   639.743 &  709.130  \nl
BSS21&  10850 &   19.326 &   18.753 &   630.984 &  656.546 &  &BSS61&  13487 &   19.892 &   19.019 &   683.171 &  635.413  \nl
BSS22&  14639 &   19.330 &   18.465 &   684.581 &  652.878 &  &BSS62&  10845 &   19.904 &   19.054 &   652.011 &  642.520  \nl
BSS23&  21847 &   19.350 &   18.872 &   518.844 &  709.890 &  &BSS63&  15094 &   19.909 &   19.214 &   679.814 &  667.246  \nl
BSS24&  43935 &   19.358 &   18.511 &   650.218 &  603.417 &  &BSS64&  15365 &   19.911 &   19.207 &   678.079 &  649.712  \nl
BSS25&  11978 &   19.362 &   18.740 &   676.616 &  664.153 &  &BSS65&  11461 &   19.926 &   18.952 &   666.185 &  654.131  \nl
BSS26&  11605 &   19.396 &   18.307 &   691.932 &  642.124 &  &BSS66&  12847 &   19.929 &   18.999 &   706.911 &  673.669  \nl
BSS27&  11663 &   19.436 &   18.951 &   684.123 &  649.454 &  &BSS67&  13302 &   19.936 &   19.411 &   679.628 &  624.777  \nl
BSS28&  14562 &   19.460 &   18.963 &   695.810 &  625.733 &  &BSS68&  12060 &   19.939 &   19.409 &   643.727 &  687.863  \nl
BSS29&  13387 &   19.481 &   18.500 &   672.694 &  635.620 &  &BSS69&  14825 &   19.946 &   19.509 &   685.224 &  633.806  \nl
BSS30&  12229 &   19.518 &   18.661 &   692.561 &  661.135 &  &BSS70&  13494 &   19.948 &   19.297 &   647.515 &  659.453  \nl
BSS31&  15257 &   19.593 &   19.019 &   676.614 &  653.890 &  &BSS71&  31752 &   19.955 &   19.140 &   563.856 &  625.901  \nl
BSS32&  43749 &   19.615 &   19.121 &   695.164 &  578.106 &  &BSS72&  13541 &   19.957 &   19.386 &   698.976 &  630.335  \nl
BSS33&  10628 &   19.616 &   18.940 &   654.840 &  633.363 &  &BSS73&  13908 &   19.958 &   19.074 &   696.672 &  665.286  \nl
BSS34&  14496 &   19.621 &   18.927 &   663.863 &  624.841 &  &BSS74&  13503 &   19.961 &   18.903 &   666.971 &  647.575  \nl
BSS35&  10357 &   19.655 &   19.206 &   665.331 &  616.820 &  &BSS75&  43763 &   19.963 &   19.155 &   657.903 &  511.669  \nl
BSS36&  15212 &   19.656 &   18.723 &   674.464 &  644.874 &  &BSS76&  15006 &   19.971 &   19.305 &   688.498 &  656.824  \nl
BSS37&  15344 &   19.678 &   19.008 &   679.951 &  645.845 &  &BSS77&  30428 &   19.971 &   19.080 &   507.443 &  657.791  \nl
BSS38&  12032 &   19.683 &   18.963 &   695.952 &  653.085 &  &BSS78&  15044 &   19.979 &   19.327 &   712.280 &  656.451  \nl
BSS39&  10703 &   19.697 &   19.028 &   663.739 &  630.209 &  &BSS79&  41134 &   19.981 &   19.311 &   636.288 &  589.634  \nl
BSS40&  12408 &   19.708 &   18.873 &   659.253 &  688.351 &  &BSS80&  21457 &   19.992 &   18.849 &   627.268 &  811.920  \nl
BSS81&  11652 &   19.996 &   19.379 &   702.232 &  636.957 &  &BSS121&  14201 &   20.116 &   19.364 &   681.917 &  632.980  \nl
BSS82&  13322 &   19.998 &   19.512 &   697.476 &  614.573 &  &BSS122&  15014 &   20.120 &   19.491 &   693.555 &  658.775  \nl
BSS83&  11941 &   19.998 &   19.349 &   699.408 &  647.989 &  &BSS123&  20203 &   20.127 &   19.190 &   624.702 &  691.713  \nl
BSS84&  21118 &   20.002 &   19.162 &   578.326 &  697.644 &  &BSS124&  15032 &   20.130 &   18.987 &   670.213 &  653.507  \nl
BSS85&  15101 &   20.006 &   19.071 &   677.816 &  646.873 &  &BSS125&  20115 &   20.135 &   19.272 &   627.499 &  690.667  \nl
BSS86&  14958 &   20.009 &   19.287 &   684.011 &  644.586 &  &BSS126&  15360 &   20.138 &   19.088 &   680.604 &  650.731  \nl
BSS87&  41803 &   20.010 &   18.849 &   668.254 &  580.809 &  &BSS127&  43938 &   20.139 &   19.358 &   648.548 &  597.091  \nl
BSS88&  12686 &   20.017 &   19.194 &   703.630 &  669.480 &  &BSS128&  15248 &   20.142 &   19.068 &   672.450 &  650.975  \nl
BSS89&  11241 &   20.018 &   19.467 &   687.897 &  632.313 &  &BSS129&  11860 &   20.148 &   19.556 &   687.557 &  653.547  \nl
BSS90&  10881 &   20.019 &   19.386 &   672.884 &  630.670 &  & &   &    &    &     &      \nl
BSS91&  15216 &   20.022 &   19.299 &   674.868 &  645.430 &  & &   &    &     &     &     \nl
BSS92&  15335 &   20.026 &   19.059 &   679.280 &  644.419 &  & &   &  {\bf Faint}  &    {\bf BSS} &   &    \nl
BSS93&  15123 &   20.030 &   19.394 &   699.489 &  656.087 &  & &   &    &     &    &      \nl
BSS94&  10879 &   20.031 &   19.123 &   647.484 &  647.048 &  &BSS130&  10389 &   20.158 &   19.610 &   642.587 &  632.806  \nl
BSS95&  30577 &   20.039 &   19.115 &   491.789 &  652.232 &  &BSS131&  13396 &   20.159 &   19.373 &   672.170 &  637.232  \nl
BSS96&  43669 &   20.042 &   19.153 &   668.734 &  589.252 &  &BSS132&  14547 &   20.164 &   19.385 &   675.904 &  633.236  \nl
BSS97&  15343 &   20.046 &   19.254 &   678.100 &  644.472 &  &BSS133&  31788 &   20.166 &   19.299 &   573.854 &  644.047  \nl
BSS98&  13763 &   20.054 &   19.442 &   646.040 &  684.170 &  &BSS134&  20040 &   20.172 &   19.294 &   674.662 &  756.305  \nl
BSS99&  14601 &   20.058 &   18.853 &   668.011 &  656.901 &  &BSS135&  14237 &   20.174 &   19.365 &   664.719 &  650.635  \nl
BSS100&  43824 &   20.059 &   19.038 &   729.577 &  526.286 &  &BSS136&  10685 &   20.182 &   19.526 &   661.909 &  630.766  \nl
BSS101&  14216 &   20.062 &   19.075 &   676.782 &  638.048 &  &BSS137&  15244 &   20.183 &   19.014 &   671.914 &  649.863  \nl
BSS102&  10376 &   20.063 &   19.553 &   634.087 &  637.975 &  &BSS138&  12870 &   20.185 &   19.517 &   714.991 &  669.596  \nl
BSS103&  14337 &   20.065 &   19.478 &   677.405 &  657.863 &  &BSS139&  13419 &   20.192 &   19.597 &   695.128 &  623.244  \nl
BSS104&  14402 &   20.068 &   19.354 &   684.148 &  666.191 &  &BSS140&  14286 &   20.192 &   19.133 &   682.808 &  643.842  \nl
BSS105&  13416 &   20.071 &   19.303 &   682.384 &  631.701 &  &BSS141&  11582 &   20.194 &   19.117 &   657.931 &  663.622  \nl
BSS106&  14321 &   20.074 &   19.359 &   677.821 &  655.064 &  &BSS142&  14828 &   20.199 &   19.385 &   672.974 &  641.798  \nl
BSS107&  13755 &   20.076 &   19.162 &   701.435 &  647.361 &  &BSS143&  15219 &   20.200 &   19.093 &   674.284 &  646.426  \nl
BSS108&  14509 &   20.077 &   19.387 &   672.673 &  626.527 &  &BSS144&  10271 &   20.205 &   19.557 &   663.338 &  614.546  \nl
BSS109&  14192 &   20.081 &   19.385 &   666.286 &  641.646 &  &BSS145&  12251 &   20.209 &   19.437 &   693.983 &  661.022  \nl
BSS110&  14312 &   20.083 &   19.041 &   674.578 &  656.175 &  &BSS146&  15269 &   20.214 &   19.434 &   682.861 &  639.683  \nl
BSS111&  11913 &   20.086 &   19.299 &   663.298 &  670.750 &  &BSS147&  12561 &   20.215 &   19.245 &   716.134 &  656.440  \nl
BSS112&  13430 &   20.087 &   19.363 &   664.277 &  644.634 &  &BSS148&  11260 &   20.215 &   18.953 &   688.767 &  632.229  \nl
BSS113&  10507 &   20.093 &   19.053 &   668.047 &  620.675 &  &BSS149&  15261 &   20.216 &   19.334 &   676.303 &  654.429  \nl
BSS114&  14332 &   20.096 &   19.092 &   683.058 &  652.677 &  &BSS150&  10364 &   20.216 &   19.536 &   643.146 &  631.635  \nl
BSS115&  40252 &   20.099 &   19.255 &   601.244 &  616.704 &  &BSS151&  14033 &   20.216 &   19.217 &   691.964 &  684.797  \nl
BSS116&  15166 &   20.104 &   18.908 &   669.725 &  654.929 &  &BSS152&  13521 &   20.219 &   19.338 &   651.292 &  659.059  \nl
BSS117&  43878 &   20.110 &   19.371 &   648.907 &  579.637 &  &BSS153&  15332 &   20.221 &   19.245 &   678.972 &  643.338  \nl
BSS118&  43042 &   20.110 &   19.024 &   625.900 &  377.749 &  &BSS154&  12812 &   20.223 &   19.399 &   690.496 &  683.165  \nl
BSS119&  13904 &   20.110 &   19.287 &   682.004 &  674.804 &  &BSS155&  12615 &   20.231 &   19.360 &   705.921 &  665.231  \nl
BSS120&  14882 &   20.114 &   19.535 &   689.140 &  649.773 &  &BSS156&  15159 &   20.235 &   19.561 &   671.020 &  646.408  \nl
BSS157&  14266 &   20.237 &   19.465 &   661.788 &  656.032 &  &BSS197&  12463 &   20.374 &   19.544 &   713.418 &  654.890  \nl
BSS158&  15186 &   20.238 &   19.124 &   696.072 &  655.066 &  &BSS198&  20317 &   20.379 &   19.196 &   602.803 &  666.827  \nl
BSS159&  13357 &   20.243 &   19.458 &   681.030 &  628.110 &  &BSS199&  13841 &   20.384 &   19.774 &   687.317 &  665.039  \nl
BSS160&  15150 &   20.245 &   19.028 &   675.549 &  641.440 &  &BSS200&  12845 &   20.385 &   19.382 &   699.523 &  678.315  \nl
BSS161&  15218 &   20.255 &   19.086 &   673.558 &  646.634 &  &BSS201&  14933 &   20.385 &   19.474 &   695.746 &  680.347  \nl
BSS162&  15239 &   20.261 &   19.528 &   671.995 &  648.257 &  &BSS202&  14284 &   20.387 &   19.453 &   682.021 &  644.108  \nl
BSS163&  14532 &   20.262 &   19.558 &   678.935 &  627.946 &  &BSS203&  14236 &   20.389 &   19.468 &   664.509 &  650.331  \nl
BSS164&  40791 &   20.266 &   19.446 &   621.682 &  596.032 &  &BSS204&  11488 &   20.390 &   19.439 &   688.841 &  640.450  \nl
BSS165&  14798 &   20.268 &   19.002 &   716.549 &  649.574 &  &BSS205&  14984 &   20.390 &   19.794 &   682.100 &  651.627  \nl
BSS166&  13310 &   20.269 &   19.516 &   693.353 &  616.340 &  &BSS206&  14267 &   20.395 &   19.456 &   661.925 &  656.421  \nl
BSS167&  13297 &   20.279 &   19.339 &   645.724 &  646.381 &  &BSS207&  20351 &   20.395 &   19.523 &   644.823 &  731.688  \nl
BSS168&  11301 &   20.284 &   19.477 &   678.524 &  640.427 &  &BSS208&  14114 &   20.402 &   19.321 &   661.224 &  629.283  \nl
BSS169&  10814 &   20.287 &   19.568 &   658.343 &  637.362 &  &BSS209&  11910 &   20.402 &   19.616 &   703.331 &  644.554  \nl
BSS170&  40934 &   20.287 &   19.395 &   631.342 &  599.438 &  &BSS210&  13690 &   20.403 &   19.590 &   699.971 &  643.245  \nl
BSS171&  40272 &   20.287 &   19.371 &   606.178 &  622.875 &  &BSS211&  13882 &   20.403 &   19.451 &   689.814 &  667.414  \nl
BSS172&  20628 &   20.294 &   19.172 &   599.222 &  685.074 &  &BSS212&  15100 &   20.405 &   19.816 &   676.426 &  647.330  \nl
BSS173&  15196 &   20.297 &   19.470 &   669.547 &  640.959 &  &BSS213&  22431 &   20.406 &   19.527 &   642.283 &  737.231  \nl
BSS174&  15132 &   20.299 &   19.572 &   661.906 &  643.559 &  &BSS214&  14127 &   20.410 &   19.772 &   655.388 &  635.092  \nl
BSS175&  14247 &   20.302 &   19.498 &   694.785 &  632.791 &  &BSS215&  12310 &   20.410 &   19.408 &   702.222 &  657.056  \nl
BSS176&  15179 &   20.306 &   19.639 &   693.309 &  653.493 &  &BSS216&  14531 &   20.413 &   19.558 &   678.742 &  627.527  \nl
BSS177&  22510 &   20.306 &   19.074 &   601.699 &  668.373 &  &BSS217&  14963 &   20.418 &   19.560 &   685.408 &  643.859  \nl
BSS178&  15098 &   20.307 &   19.232 &   677.015 &  647.189 &  &BSS218&  12422 &   20.418 &   19.718 &   712.591 &  654.067  \nl
BSS179&  30372 &   20.310 &   19.582 &   562.428 &  628.251 &  &BSS219&  15108 &   20.422 &   19.240 &   686.117 &  645.326  \nl
BSS180&  14617 &   20.311 &   19.286 &   678.432 &  651.663 &  &BSS220&  15153 &   20.423 &   19.630 &   671.364 &  644.742  \nl
BSS181&  14572 &   20.314 &   19.561 &   662.892 &  649.512 &  &BSS221&  14652 &   20.425 &   19.322 &   665.202 &  667.023  \nl
BSS182&  13506 &   20.315 &   19.037 &   674.174 &  643.346 &  &BSS222&  14270 &   20.426 &   19.410 &   703.610 &  628.765  \nl
BSS183&  14162 &   20.330 &   19.509 &   671.462 &  631.488 &  &BSS223&  12352 &   20.428 &   19.535 &   681.541 &  671.892  \nl
BSS184&  13885 &   20.336 &   19.345 &   681.901 &  672.537 &  &BSS224&  12857 &   20.429 &   19.494 &   708.919 &  672.940  \nl
BSS185&  11964 &   20.338 &   19.497 &   710.443 &  641.391 &  &BSS225&  11408 &   20.430 &   19.575 &   648.791 &  663.522  \nl
BSS186&  10588 &   20.338 &   19.241 &   671.764 &  621.222 &  &BSS226&  13309 &   20.433 &   19.354 &   693.686 &  616.011  \nl
BSS187&  14285 &   20.344 &   19.196 &   682.435 &  643.755 &  &BSS227&  11052 &   20.433 &   19.487 &   667.882 &  639.291  \nl
BSS188&  15185 &   20.346 &   19.546 &   695.404 &  655.128 &  &BSS228&  12266 &   20.437 &   19.282 &   678.006 &  671.974  \nl
BSS189&  14252 &   20.348 &   19.504 &   663.878 &  653.352 &  &BSS229&  12940 &   20.438 &   19.528 &   728.299 &  663.606  \nl
BSS190&  13050 &   20.350 &   19.575 &   666.803 &  708.402 &  &BSS230&  13584 &   20.438 &   19.579 &   652.013 &  665.324  \nl
BSS191&  20190 &   20.352 &   19.504 &   634.013 &  704.955 &  &BSS231&  12443 &   20.439 &   19.614 &   666.583 &  684.730  \nl
BSS192&  14350 &   20.353 &   19.414 &   677.232 &  660.442 &  &BSS232&  14881 &   20.445 &   19.952 &   688.595 &  649.890  \nl
BSS193&  14150 &   20.355 &   19.417 &   656.920 &  639.568 &  &BSS233&  12444 &   20.445 &   19.331 &   672.576 &  680.794  \nl
BSS194&  15195 &   20.367 &   19.231 &   671.005 &  639.621 &  &BSS234&  15104 &   20.449 &   19.951 &   676.364 &  647.734  \nl
BSS195&  12934 &   20.368 &   19.620 &   728.742 &  663.058 &  &BSS235&  14896 &   20.450 &   19.622 &   688.967 &  654.023  \nl
BSS196&  15334 &   20.373 &   19.449 &   679.617 &  643.729 &  &BSS236&  14362 &   20.450 &   19.307 &   701.938 &  645.824  \nl
BSS237&  11556 &   20.453 &   19.384 &   662.948 &  659.486 &  &BSS277&  13930 &   20.524 &   20.006 &   715.254 &  654.985  \nl
BSS238&  14605 &   20.454 &   19.547 &   671.446 &  654.854 &  &BSS278&  13420 &   20.525 &   19.524 &   695.694 &  623.308  \nl
BSS239&  14320 &   20.457 &   19.387 &   678.262 &  654.542 &  &BSS279&  10991 &   20.525 &   19.838 &   632.049 &  660.775  \nl
BSS240&  13569 &   20.458 &   19.483 &   686.270 &  641.558 &  &BSS280&  13289 &   20.526 &   19.440 &   654.362 &  640.406  \nl
BSS241&  13325 &   20.458 &   19.706 &   661.517 &  637.933 &  &BSS281&  13242 &   20.527 &   19.555 &   650.050 &  636.474  \nl
BSS242&  15214 &   20.459 &   19.562 &   674.919 &  645.062 &  &BSS282&  13444 &   20.528 &   19.865 &   652.815 &  652.799  \nl
BSS243&  20828 &   20.461 &   19.456 &   586.091 &  679.750 &  &BSS283&  14982 &   20.528 &   19.929 &   683.080 &  651.984  \nl
BSS244&  14972 &   20.463 &   19.718 &   682.140 &  649.020 &  &BSS284&  10891 &   20.529 &   19.946 &   637.064 &  654.441  \nl
BSS245&  15004 &   20.463 &   19.911 &   690.168 &  655.668 &  &BSS285&  10603 &   20.530 &   19.548 &   648.358 &  636.828  \nl
BSS246&  13797 &   20.464 &   19.827 &   697.571 &  653.898 &  &BSS286&  13577 &   20.531 &   19.322 &   695.771 &  635.856  \nl
BSS247&  14382 &   20.464 &   19.809 &   677.688 &  665.821 &  &BSS287&  15319 &   20.531 &   19.547 &   677.150 &  649.156  \nl
BSS248&  13336 &   20.466 &   19.608 &   689.062 &  621.485 &  &BSS288&  13629 &   20.531 &   19.670 &   691.106 &  643.982  \nl
BSS249&  13654 &   20.471 &   19.381 &   689.822 &  647.018 &  &BSS289&  41514 &   20.532 &   19.634 &   652.845 &  583.432  \nl
BSS250&  13229 &   20.474 &   19.487 &   676.311 &  617.896 &  &BSS290&  13009 &   20.532 &   19.594 &   714.279 &  675.590  \nl
BSS251&  14401 &   20.480 &   19.525 &   686.165 &  664.941 &  &BSS291&  14526 &   20.534 &   19.719 &   670.403 &  632.016  \nl
BSS252&  20733 &   20.482 &   19.525 &   612.826 &  711.935 &  &BSS292&  11902 &   20.535 &   19.591 &   680.951 &  659.115  \nl
BSS253&  30220 &   20.483 &   19.695 &   476.277 &  701.749 &  &BSS293&  14707 &   20.536 &   19.707 &   671.712 &  614.140  \nl
BSS254&  10565 &   20.483 &   19.442 &   659.843 &  627.875 &  &BSS294&  10414 &   20.539 &   19.564 &   672.335 &  614.438  \nl
BSS255&  14884 &   20.489 &   19.943 &   690.085 &  649.827 &  &BSS295&  12438 &   20.540 &   19.761 &   702.380 &  661.270  \nl
BSS256&  11911 &   20.491 &   19.248 &   708.752 &  641.008 &  &BSS296&  11866 &   20.541 &   19.600 &   645.267 &  681.357  \nl
BSS257&  15102 &   20.491 &   19.566 &   678.047 &  647.186 &  &BSS297&  13982 &   20.543 &   19.706 &   679.266 &  685.562  \nl
BSS258&  10931 &   20.494 &   19.547 &   649.233 &  647.593 &  &BSS298&  13417 &   20.543 &   19.868 &   670.957 &  638.909  \nl
BSS259&  14599 &   20.496 &   19.602 &   676.628 &  651.761 &  &BSS299&  14373 &   20.545 &   19.790 &   668.787 &  670.565  \nl
BSS260&  14942 &   20.498 &   19.792 &   672.602 &  637.606 &  &BSS300&  12985 &   20.547 &   19.862 &   708.904 &  678.388  \nl
BSS261&  12419 &   20.498 &   19.975 &   699.785 &  662.194 &  &BSS301&  10150 &   20.547 &   19.938 &   626.381 &  633.368  \nl
BSS262&  14552 &   20.501 &   19.459 &   669.193 &  638.541 &  &BSS302&  12825 &   20.548 &   19.409 &   686.603 &  685.971  \nl
BSS263&  13015 &   20.503 &   19.988 &   693.784 &  689.630 &  &BSS303&  13263 &   20.548 &   19.605 &   672.373 &  625.280  \nl
BSS264&  14748 &   20.506 &   19.680 &   674.291 &  651.745 &  &BSS304&  12134 &   20.549 &   19.894 &   713.490 &  644.367  \nl
BSS265&  12217 &   20.507 &   19.623 &   702.089 &  654.429 &  &BSS305&  10638 &   20.550 &   19.803 &   693.035 &  608.847  \nl
BSS266&  14993 &   20.510 &   19.616 &   686.722 &  656.283 &  & &       &      &      &       &       \nl
BSS267&  11223 &   20.513 &   19.905 &   693.393 &  628.203 &  & &       &      &      &       &       \nl
BSS268&  11552 &   20.513 &   19.971 &   700.215 &  635.018 &  & &       &      &      &       &       \nl
BSS269&  11943 &   20.516 &   19.672 &   676.643 &  663.071 &  & &       &      &      &       &       \nl
BSS270&  10806 &   20.517 &   19.400 &   649.510 &  642.744 &  & &       &      &      &       &       \nl
BSS271&  10260 &   20.517 &   19.949 &   669.445 &  610.096 &  & &       &      &      &       &       \nl
BSS272&  14276 &   20.517 &   19.396 &   675.725 &  648.328 &  & &       &      &      &       &       \nl
BSS273&  13664 &   20.518 &   19.921 &   688.554 &  649.185 &  & &       &      &      &       &       \nl
BSS274&  12380 &   20.518 &   19.366 &   706.500 &  656.701 &  & &       &      &      &       &       \nl
BSS275&  13413 &   20.521 &   19.664 &   674.422 &  636.487 &  & &       &      &      &       &       \nl
BSS276&  11715 &   20.524 &   19.434 &   692.627 &  645.562 &  & &       &      &      &       &       \nl
\enddata
\end{deluxetable}

\newpage
\hoffset = 0mm
\begin{deluxetable}{cccccc}
\tablewidth{15truecm}
\label{lmebss}
\tablecaption{The E-BSS population in M80.}
\tablehead{
\colhead{ $Name$} &
\colhead{ $Id$} &
\colhead{$V$} &
\colhead{$U$} &
\colhead{$X$} &
\colhead{$Y$} }
\startdata
E-BSS1&14764 &  15.053  & 16.508 &  668.291 &   660.010\nl
E-BSS2&14945 &  15.589  & 16.532 &  673.819  &  638.034\nl
E-BSS3&15346 &  15.171  & 16.558 &  681.072 &   642.964\nl
E-BSS4&15350 &  15.580  & 16.559 &  679.837  &  649.538\nl
E-BSS5&20898 &  15.761  & 16.607 &  628.347  &  747.758\nl
E-BSS6&15300 &  15.289  & 16.611 &  679.280 &   641.327\nl
E-BSS7&13422 &  15.318  & 16.663 &  687.622 &   629.165\nl
E-BSS8&15333 &  15.674  & 16.665 &  679.198  &  643.797\nl
E-BSS9&15273 &  15.199  & 16.665 &  683.524 &   640.650\nl
E-BSS10&13891 &  15.327  & 16.690 &  693.230 &   665.446\nl
E-BSS11&14097 &  15.406  & 16.698 &  670.429 &   620.216\nl
E-BSS12&14710 &  15.359  & 16.713 &  647.164 &   633.821\nl
E-BSS13&30914 &  15.360 &  16.727 &  481.321 &   614.829\nl
E-BSS14&13121 &  15.490  & 16.741 &  656.676 &   612.601\nl
E-BSS15&21514 &  15.321  & 16.745 &  542.077 &   692.453\nl 
E-BSS16&13247 &  15.704  & 16.759 &  638.376  &  645.053\nl
E-BSS17&13137 &  15.686  & 16.765 &  633.291  &  630.344\nl
E-BSS18&13220 &  15.843  & 16.767 &  646.076  &  636.747\nl
E-BSS19&43519 &  15.501 &  16.790 &  630.069 &   622.421\nl
\enddata
\end{deluxetable}

\clearpage
\centerline {FIGURE CAPTIONS}

{\bf Fig1} A $V$-band WFPC2 image
({\it in the background})
of M80 (negative greyscale)
and ({\it in foreground})
a zoomed map (in the UV-F255W filter) of the cluster's central
$10\arcsec\times10\arcsec$ region. The red-colored stars indicate the
identified Blue Stragglers. The blue circle has
a radius of $6\farcs 5$ from
the cluster center of gravity (which is indicated by a heavy blue
`+').
It corresponds to the core radius of the cluster ($r_c$)
as determined by this study.
 The brightest objects  in the zoomed image are horizontal branch (HB)
stars. Both the HB and BSS are easily identified in contrast to the
$V$ image which is dominated by the red giants.

\bigskip

{\bf Fig2}The $(m_{255},m_{255}-U)$-CMD
for the central region of M80.  In the left panel the whole CMD is shown.
The solid line corresponds to $U=21$.
The right panel shows the zoomed CMD in the BSS region. BSS are plotted as
solid circles. The heavy horizontal line at
$m_{255}=20.55$ corresponds to the assumed limiting magnitude for the
selected BSS. 

\bigskip

{\bf Fig3} Cumulative radial distribution ($\phi$) for BSS
(heavy solid line), E-BSS (dotted line)
compared to the HB+RGB stars (dashed line) as a function
of their projected distance ($r$) from the cluster center.

\bigskip

{\bf Fig4} The observed radial density profile (filled circles) with
respect to the center of gravity. The dashed line in Panel~b is the
best fitting King Model with $r_c=6\farcs 5$ and $c=2.0$.

 \bigskip

{\bf Fig4}Zoomed $(U,~U-V)-$CMD
of the Horizontal branch region. The E-BSS candidates
are plotted as large filled circles.

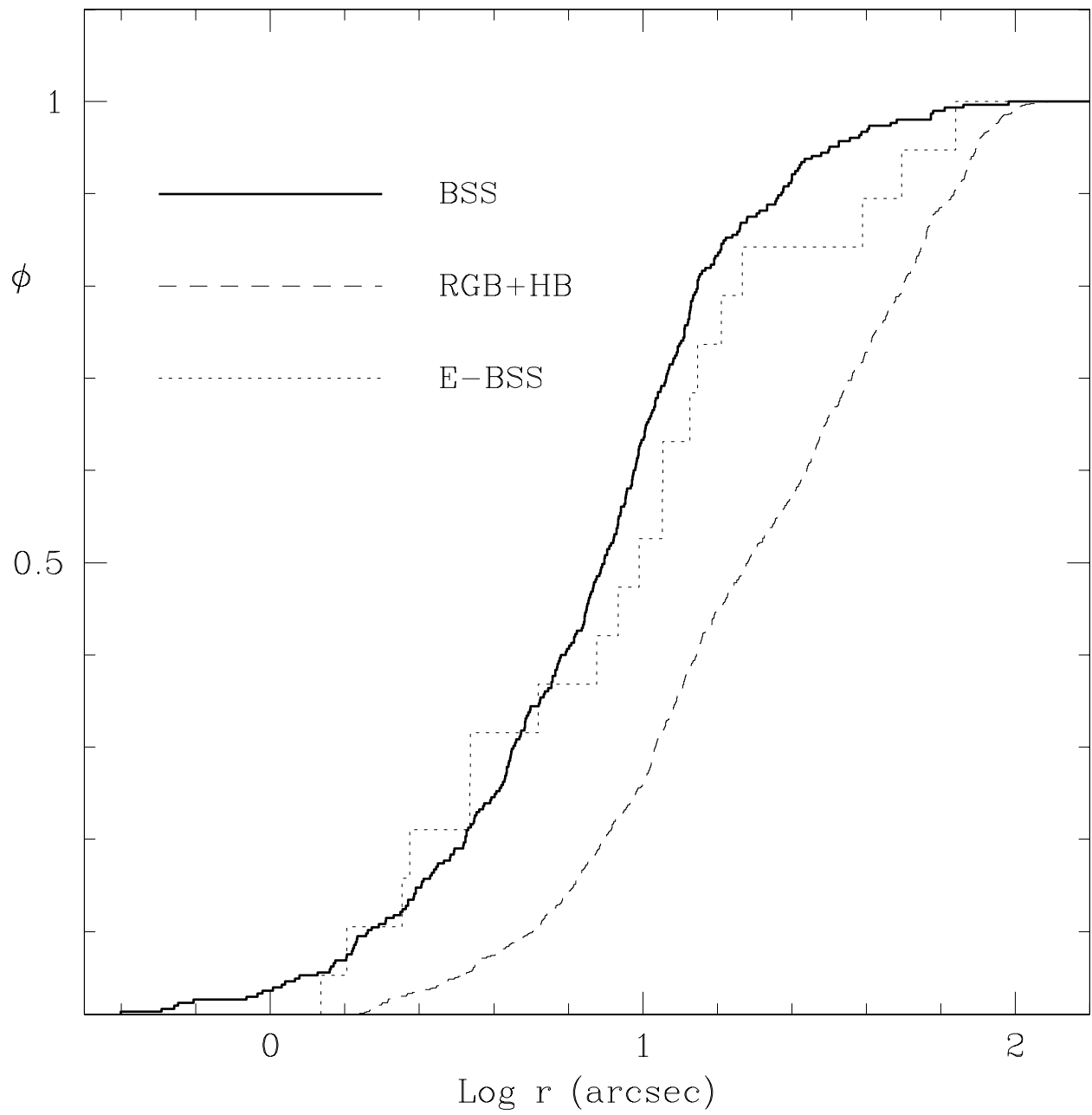
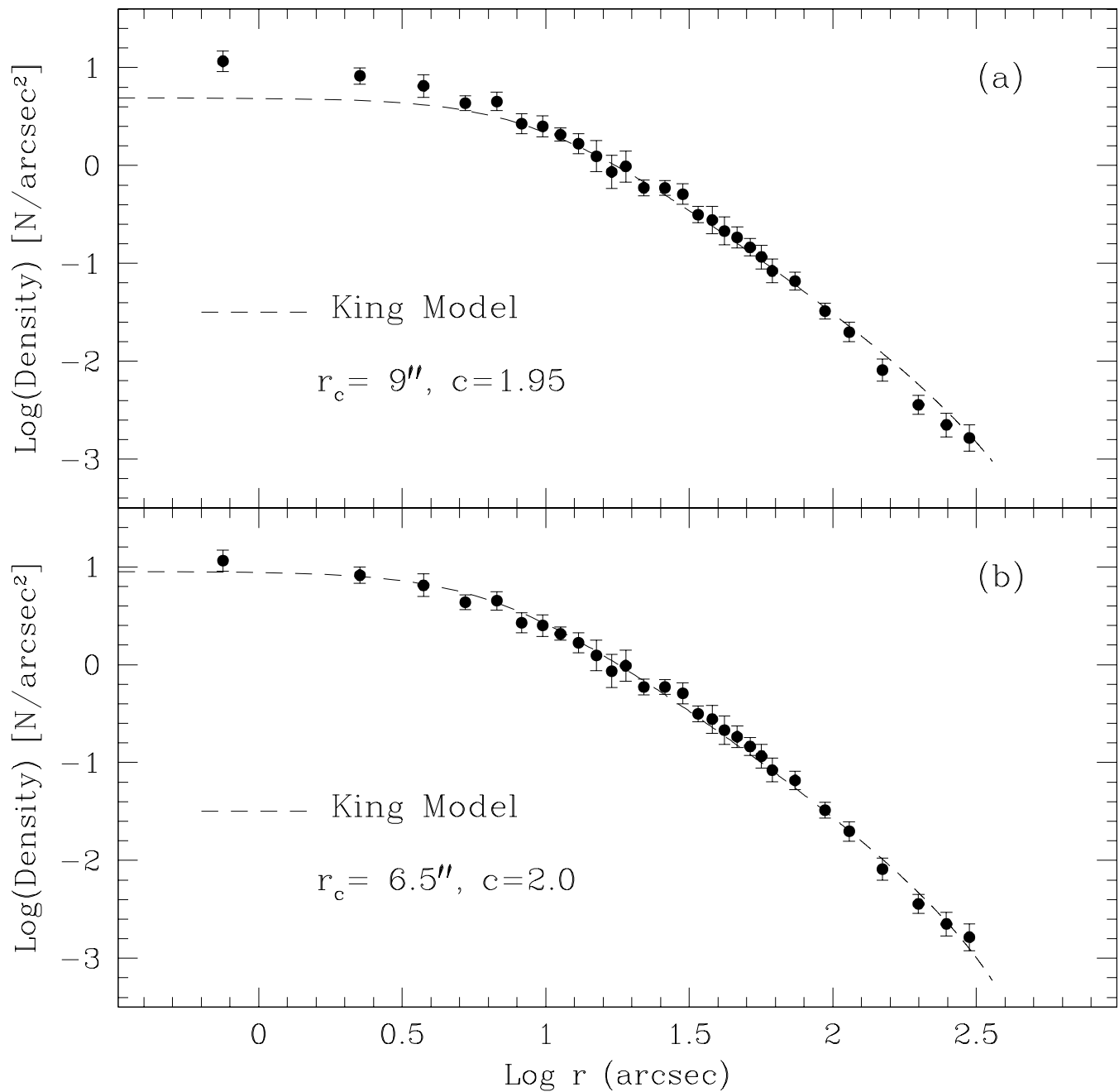
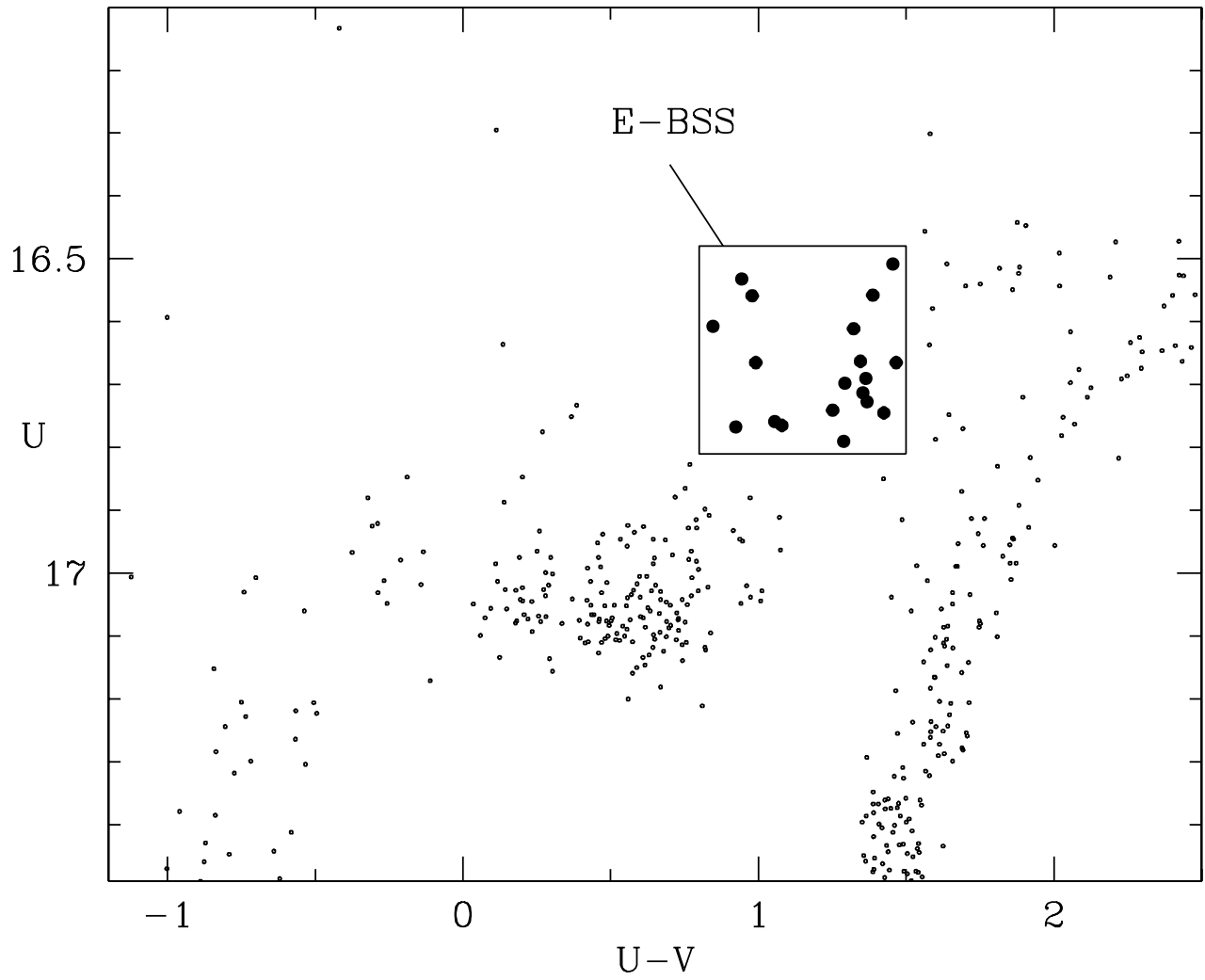
\end{document}